\documentclass[aps,prd,reprint,nofootinbib,amsmath,amssymb,amsfonts,floatfix]{revtex4-2}

\usepackage{amsmath}
\usepackage{amssymb}
\usepackage{graphicx}
\usepackage{hyperref}
\usepackage{braket}
\usepackage{geometry}
\usepackage[bottom]{footmisc}

\begin{document}

\title{Hydrodynamic Analog of the Klein Paradox: Vacuum Instability and Pair Production in a Linear Elastic Medium}

\author{Alan F. Tinoco V\'azquez}\thanks{\textbf{tinoco.v.alanfermin@gmail.com}}
\affiliation{Independent Researcher, Guadalajara, Jalisco, M\'exico}

\begin{abstract}
The Klein Paradox—the anomalous scattering of relativistic fermions off a high potential step—signals the limit of the single-particle interpretation of the Dirac equation. While Quantum Field Theory (QFT) resolves this via pair production, the microscopic mechanism is often obscured by abstract formalism. In this work, we investigate this phenomenon through the framework of Analog Gravity and Condensed Matter Physics. We utilize a hydrodynamic model wherein a relativistic particle is treated as a localized elastic excitation (defect) within a continuous linear medium. We demonstrate that when the external stress (potential) exceeds the medium’s binding energy threshold ($V > 2mc^2$), the system undergoes a mechanical instability analogous to dielectric breakdown. This instability naturally generates modes with inverted topological winding, which we identify as antiparticles. By solving the boundary conditions for this elastic system, we reproduce the transmission coefficients of Hansen and Ravndal and recover the Schwinger limit for pair production rates. This approach provides a clear pedagogical model based on continuum mechanics to visualize vacuum decay processes, suggesting that the "paradox" is simply the elastic response of a medium under supercritical stress. This mechanical analogy serves as a pedagogical bridge for graduate students in condensed matter physics and advanced materials science, offering a concrete visualization of vacuum instability that complements standard abstract QFT derivations.
\newline
\textbf{Eur. J. Phys. Accepted Manuscript (2026); DOI:} \href{https://iopscience.iop.org/article/10.1088/1361-6404/ae5903}{10.1088/1361-6404/ae5903} \footnote{This is the Author's Accepted Manuscript. The final version of record has been published in European Journal of Physics and is available at DOI: \href{https://iopscience.iop.org/article/10.1088/1361-6404/ae5903}{10.1088/1361-6404/ae5903}. This manuscript is available for reuse under a CC BY-NC-ND licence after the 12-month embargo period.}
\end{abstract}

\maketitle

\section{Introduction}

Since its formulation in 1929, the Klein Paradox \cite{Klein1929} has served as a critical test case for the interpretation of relativistic quantum mechanics. When a Dirac electron encounters a potential step of height $V > E + mc^2$, the single-particle Dirac equation predicts a reflection coefficient $R > 1$ and a non-vanishing transmission probability into states of formally negative kinetic energy.

As detailed in standard texts like Greiner \cite{Greiner}, this result is paradoxical only if one insists on conserving the probability density of a single particle. Greiner explicitly calculates the currents to show that the reflected flux exceeds the incident flux, necessitating a many-body interpretation, where the potential stimulates the creation of particle-antiparticle pairs. Similarly, Itzykson and Zuber \cite{Itzykson} use this breakdown to motivate the transition to Quantum Field Theory (QFT), arguing that the Dirac equation in strong fields describes charge redistribution in the vacuum rather than single-particle trajectories.

While the QFT resolution is mathematically robust, the physical intuition often remains opaque. Modern treatments connect the paradox to the Schwinger Effect \cite{Schwartz, Xu2015}---the creation of pairs in a strong electric field---but usually treat the "vacuum" as an abstract state subject to operator creation/annihilation rules.

While traditionally a subject of high-energy physics, the Klein Paradox has found renewed relevance in condensed matter systems. The realization of massless Dirac fermions in graphene and topological insulators has allowed for the experimental observation of Klein tunneling \cite{Katsnelson2006, Stander2009}. Furthermore, recent experimental breakthroughs in phononic crystals \cite{Jiang2020, Wu2022} have demonstrated this phenomenon in macroscopic acoustic media, validating the hydrodynamic analogy utilized in this work, making the intuitive understanding of this phenomenon crucial for modern materials science.

Recent pedagogical works have utilized multiple scattering expansions to visualize the time-dependent dynamics of the paradox for scalar particles \cite{Alkhateeb2022}. In this work, we take a complementary approach through the framework of \textit{Analog Gravity} and \textit{Condensed Matter Physics} \cite{Volovik, Unruh}. We explore a hydrodynamic toy model wherein the relativistic fermion is treated as a localized elastic excitation (defect) within a continuous linear medium. In this analog, "mass" corresponds to a resonant frequency gap, and the spinor structure arises from the internal vibrational degrees of freedom required for stability.

We demonstrate that when the external stress (potential) exceeds the medium's binding energy threshold ($V > 2mc^2$), the system undergoes a mechanical instability analogous to \textit{dielectric breakdown}. This instability naturally generates modes with inverted topological winding, which we identify as antiparticles. By solving the boundary conditions for this elastic system, we reproduce the transmission coefficients of Hansen and Ravndal \cite{Hansen1981}, suggesting that the "paradoxical" reflection can be understood as the elastic response of a medium under supercritical stress.

The novelty of this work lies not in calculating new scattering coefficients—which necessarily match standard QFT predictions—but in providing a \textbf{deterministic, mechanical mechanism} for the pair creation process. While standard theory treats the vacuum instability as a probabilistic transition governed by operator algebra, our hydrodynamic framework reveals the underlying kinematics of the 'breakdown'. It offers a concrete physical picture where the 'mysterious' appearance of antiparticles is demystified as the natural relaxation of a medium under supercritical stress, bridging the conceptual gap between single-particle quantum mechanics and many-body field theory.

\subsection*{Pedagogical Aim:}

Standard instruction on the Klein Paradox typically aims to demonstrate the limits of single-particle quantum mechanics and motivate relativistic Quantum Field Theory (QFT) in a high-energy physics context. However, with the advent of Dirac materials such as graphene and topological insulators, the Klein Paradox has become a distinctly observable phenomenon in solid-state physics. 

Consequently, this article is specifically addressed to graduate students and instructors in courses dedicated to \textit{Quantum Field Theory for Condensed Matter Physics} or \textit{Advanced Solid State Physics}. In these contexts, concepts such as elastic defects, topological anchoring, and continuous media are standard prerequisites. By leveraging the framework of Analog Gravity, instructors can use this hydrodynamic model to:

\begin{enumerate}
   \item Provide a tangible, mechanical intuition for Dirac fermions emerging from lattice/medium dynamics.
    \item Visualize "pair creation" and vacuum instability not merely as abstract operator algebra, but as a deterministic dielectric-like breakdown of a continuous medium under stress.
    \item Bridge the conceptual gap between high-energy QFT phenomena and their macroscopic analogs in condensed matter systems.
\end{enumerate}
\textbf{Section \ref{sec:pedagogy}} provides specific guidelines, a comparative table, and suggested exercises for instructors to implement this hydrodynamic model. By grounding the mathematics of Dirac fermions in continuum mechanics, this approach serves as an intuitive prelude to advanced topics in topological matter and analog gravity.

\section{The Elastic Defect Ansatz}

To construct a hydrodynamic analog for the Dirac field, we adopt a model inspired by the hydrodynamics of defects in condensed matter \cite{Volovik}. We assume the background vacuum behaves as a continuous linear medium characterized by a wave propagation speed $c$ and a restorative stiffness parameter $\kappa$, as illustrated in \textbf{Fig. \ref{fig:model}}.

\subsection{Dispersion and Mass}
A localized defect in this medium is not a rigid body but a standing wave pattern. Its stability requires a resonance with the background, leading to the dispersion relation for a massive field:
\begin{equation}
    \omega^2 = c^2 k^2 + \omega_0^2
\end{equation}
Here, $\omega_0$ is the characteristic cutoff frequency of the medium's response to the defect. Following the de Broglie relations $E = \hbar\omega$ and $p = \hbar k$, we identify the rest mass energy. Similar to the energy gap in a superconductor or the effective mass of a phonon, this mass is defined as the energy cost of the confinement resonance:
\begin{equation}
    mc^2 \equiv \hbar \omega_0
\end{equation}

\subsection{Internal Structure and The Composite Spinor}
As detailed in the derivation provided in \textbf{Appendix \ref{sec:app}}, the requirement for a causal, first-order evolution of such a wave-defect in a relativistic medium necessitates a multicomponent state vector. Within this elastic framework, the state space arises as the tensor product of two independent binary physical properties:
\begin{enumerate}
    \item \textbf{Vibrational Polarization ($V_{\text{spin}}$):} The two orthogonal modes of transverse vibration relative to the propagation axis.
    \item \textbf{Topological Anchoring ($V_{\text{anchor}}$):} The orientation of the defect's connection to the vacuum medium (analogous to charge).
\end{enumerate}
The resulting state function is a 4-component spinor $\Psi \in V_{\text{spin}} \otimes V_{\text{anchor}}$.
Physically, the Dirac matrices map directly to operations on these subspaces. Crucially for the Klein paradox, the operator $\beta$ acts as the \textit{topological index} on $V_{\text{anchor}}$. Its eigenvalues $\pm 1$ distinguish between the ``normal'' anchoring of the defect (matter) and the ``inverted'' anchoring (antimatter). The mass term $\beta mc^2$ implies that the confinement energy is an intrinsic property of this anchoring stability.

\begin{figure*}[t]
    \centering
    \includegraphics[width=0.8\textwidth]{Figure1_Model.png} 
    \caption{\textbf{The Hydrodynamic Analog of Mass.} (a) Schematic representation of the relativistic particle as a localized standing wave defect within a continuous elastic medium. (b) The dispersion relation of the medium. The rest mass corresponds to the cutoff frequency $\omega_0$, physically interpreted as the minimum vibrational energy required to sustain the defect against the medium's restorative stiffness. Modes below this gap are evanescent.}
    \label{fig:model}
\end{figure*}

\section{The Klein Potential as Mechanical Stress}

We now consider the defect interacting with an external step potential $V(z)$:
\begin{equation}
    V(z) = \begin{cases} 
    0 & z < 0 \quad (\text{Region I}) \\
    V_0 & z > 0 \quad (\text{Region II})
    \end{cases}
\end{equation}
In this mechanical model, $V(z)$ represents a shift in the local energy density of the medium (or an external tension). The total Hamiltonian is \footnote{Note: Mathematically, $V(z)$ enters as the time-component of a vector potential, representing an external energy bias applied to the medium, distinct from a scalar mass perturbation}:
\begin{equation}
    H = c \alpha_z p_z + \beta mc^2 + V(z)
\end{equation}

We seek stationary states of the form $\Psi(z,t) = \psi(z)e^{-iEt/\hbar}$ with energy $E > 0$ incident from the left.

\subsection{Region I (Incident + Reflected)}
In the stress-free region ($z<0$), the dispersion relation is standard: $E^2 = c^2 p_1^2 + m^2 c^4$. The wave number is $k_1 = \sqrt{E^2 - m^2c^4}/\hbar c$. The solution is a superposition of an incident wave traveling right and a reflected wave traveling left:
\begin{equation}
    \psi_{I}(z) = u(p_1) e^{ik_1 z} + R u(-p_1) e^{-ik_1 z}
\end{equation}
where $u(p)$ are the standard free spinors, $p_1 = \hbar k_1$ is the incident momentum in Region I, and $R$ is the complex reflection amplitude.

\subsection{Region II (The Supercritical Stress)}
In the potential region ($z>0$), the energy equation becomes:
\begin{equation}
    (E - V_0)^2 = c^2 p_2^2 + m^2 c^4
\end{equation}
where $p_2$ is the momentum of the wave propagating in Region II.

We are interested in the \textbf{Klein regime} (or strong field limit \cite{Holstein1998}), where the applied tension constitutes a \textit{supercritical stress}: a potential sufficiently strong to bridge the entire mass gap, defined mathematically as:
\begin{equation}
    V_0 > E + mc^2
\end{equation}

Under this condition, $(E - V_0)^2$ is large and positive, implying that the momentum $p_2$ (and wave number $k_2$) is \textit{real}. The wave propagates inside the barrier.

However, we must determine the direction of propagation. The group velocity is given by:
\begin{equation}
    v_g = \frac{\partial E}{\partial p_2} = c^2 \frac{p_2}{E - V_0}
\end{equation}
Since $p_2 = \hbar k_2$ and $V_0 > E$, the denominator is negative. For the group velocity to be positive (energy flowing away from the boundary, $v_g > 0$), we must choose the wave number $k_2$ to be \textbf{negative}.

\subsection{Structural Interpretation of Negative Modes}
In standard theory, this solution corresponds to a "negative energy electron." In our Defect Model, the interpretation is structural.
The condition $V_0 > 2mc^2$ implies that the external stress exceeds the elastic limit required to maintain the defect's normal topology. The dispersion relation allows a new mode: a defect with \textbf{inverted topological anchoring} (see \textbf{Appendix \ref{sec:app}}).
In this regime, the transmitted wave $\psi_{II}$ describes a flux of these inverted defects (antiparticles).

\section{Vacuum Instability and Flux Conservation}

We solve the continuity of the boundary condition $\psi_I(0) = \psi_{II}(0)$. Following the explicit calculations found in Greiner \cite{Greiner} and adapted to our notation, the transmission ($T$) and reflection ($R$) amplitudes yield the current conservation relation:
\begin{equation}
    J_{\text{inc}} + J_{\text{ref}} = J_{\text{trans}}
\end{equation}
However, due to the coupling to the inverted mode (negative group velocity relative to momentum), the transmitted current $J_{\text{trans}}$ flows in the opposite direction to the particle current. Normalizing the incident flux to 1, the conservation of current ($J = \psi^\dagger \alpha_z \psi$) requires:
\begin{equation}
    1 - |R|^2 = r |T|^2
\end{equation}
where $r$ is the kinematic ratio of the transmitted to incident group velocities:
\begin{equation}
    r = \frac{v_{g,II}}{v_{g,I}} = \frac{k_2 (E)}{k_1 (E - V_0)}
\end{equation}
In the Klein regime ($V_0 > E + mc^2$), causality requires the transmitted wave packet to propagate away from the boundary, implying a positive group velocity ($v_g > 0$). However, a defining feature of the Dirac equation in the supercritical region is that the probability current $J$ for the antiparticle branch flows \textit{opposite} to the group velocity \cite{Greiner, Hansen1981}. In our defect model, this occurs because the transmitted mode possesses an inverted topological anchoring. Consequently, the kinematic ratio of the transmitted to incident currents is proportional to the transmission probability $|T|^2$ weighted by the negative group velocity ratio $-\kappa$:
\begin{equation}
    \frac{J_{\text{trans}}}{J_{\text{inc}}} = -\kappa |T|^2 \quad (\text{where } \kappa > 0)
\end{equation}
Substituting this directly into the conservation law ($1 - |R|^2 = J_{\text{trans}}/J_{\text{inc}}$), we seamlessly obtain:
\begin{equation}
    1 - |R|^2 = -\kappa |T|^2 \implies |R|^2 = 1 + \kappa |T|^2
\end{equation}
This explicitly demonstrates that the reflection coefficient exceeds unity. Physically, the negative transmitted current signifies a flux of inverted defects (antiparticles) draining into the potential, which the conservation of topological charge registers as an enhancement of the reflected particle flux.

\subsection{Mechanical Breakdown vs. Probability}
\begin{figure*}[t]
    \centering
    \includegraphics[width=0.8\textwidth]{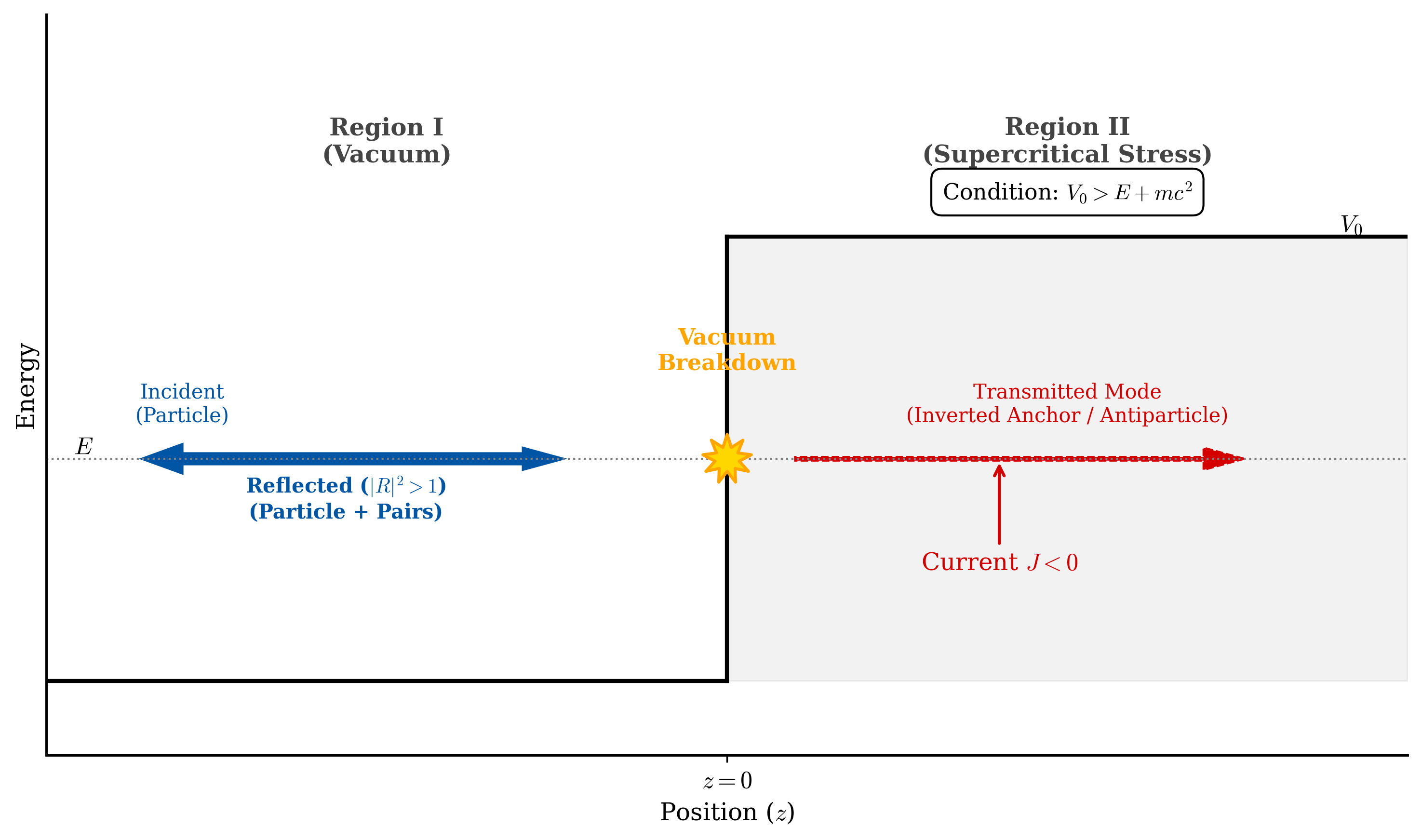} 
    \caption{\textbf{Mechanical Resolution of the Klein Paradox.} When the external stress $V_0$ exceeds the vacuum stability gap $2mc^2$, the medium undergoes mechanical breakdown at the interface. The transmitted mode (red) propagates to the right but carries a negative probability current due to its inverted topological anchoring (antiparticle nature). Conservation of flux requires the reflected particle current (blue) to exceed the incident one ($|R|^2 > 1$), physically representing the creation of defect pairs to relax the stress.}
    \label{fig:klein}
\end{figure*}

Within a single-particle probabilistic framework, $|R|^2 > 1$ is a violation of unitarity. However, in our hydrodynamic model, the interface at $z=0$ acts as an active \textbf{source} of pair production. Standard single-particle probability is not conserved; rather, it is the \textbf{topological charge} (or total energy flux) that remains balanced. The result $|R|^2 > 1$ indicates that the medium at the boundary is undergoing mechanical breakdown. The work done by the external stress $V_0$ on the vacuum medium generates the additional reflected flux by ripping the medium into pairs of defects (see \textbf{Fig. \ref{fig:klein}}).
\begin{itemize}
    \item The reflected wave consists of the original defect \textit{plus} the newly created defects (same topology).
    \item The transmitted wave consists of the partner defects with inverted topology (antiparticles) draining into the potential.
\end{itemize}
This exactly matches the "many-body" interpretation of Hansen and Ravndal \cite{Hansen1981}, who derived the pair production rate $\bar{n} = |T|^2$ using Bogoliubov transformations. Here, the "Bogoliubov transformation" is physically realized as the stress-induced mixing between the normal and inverted anchoring modes of the defect.

\section{Discussion: The Schwinger Limit}
\label{sec:5}

While the step potential is an idealization, the physical mechanism—instability due to supercritical stress—is robust. As discussed by Schwartz \cite{Schwartz} in the context of the Schwinger Effect, a constant electric field $\mathcal{E}$ produces pairs at a rate proportional to $\exp(-\pi m^2 c^3 / e\hbar \mathcal{E})$. In our hydrodynamic model, the electric field corresponds to a macroscopic mechanical strain gradient, $\nabla V$. For mechanical breakdown to occur, the work done by this external stress over the characteristic size of the defect---the Compton wavelength $\lambda_c = \hbar/mc$---must overcome the confinement energy gap $2mc^2$. Therefore, the critical strain threshold is $\nabla V_{\text{crit}} \sim 2mc^2/\lambda_c$. 

If the applied gradient is subcritical (adiabatic), the defect adjusts its frequency without breaking. The probability of tunneling through this stability gap is governed by the WKB approximation, yielding an exponential suppression factor $\exp(-\pi m^2 c^3 / \hbar \nabla V)$. Thus, the exponential suppression arises mechanically because the energy gained over a Compton wavelength is insufficient to cause instantaneous breakdown. The Klein Paradox is simply the limit of this Schwinger Effect for an infinitely sharp stress gradient ($\nabla V \to \infty$), representing the sudden, catastrophic failure of the vacuum elasticity.

\section{Pedagogical Implementation: Bridging the Conceptual Gap}
\label{sec:pedagogy}

The primary educational value of this hydrodynamic model lies in its ability to convert abstract algebraic paradoxes into concrete mechanical problems. We suggest incorporating this material into graduate courses such as \textit{Advanced Solid State Physics} or \textit{Quantum Field Theory for Condensed Matter}, where students already possess a foundational understanding of continuum mechanics and topological defects.

\subsection{Teaching Strategy: Comparative Analysis}
Instructors can use this model to address the "cognitive dissonance" students often face when transitioning from single-particle quantum mechanics to the many-body physics required by Dirac materials. By presenting the standard High-Energy QFT explanation alongside the Condensed Matter hydrodynamic analog, the physical intuition becomes tangible. We propose using \textbf{Table \ref{tab:comparison}} to guide classroom discussion on how relativistic phenomena map onto lattice/medium dynamics.

\begin{table}[h]
\centering
\caption{Comparison between Standard Dirac Theory and the Hydrodynamic Analog, useful for classroom discussion.}
\label{tab:comparison}
\begin{tabular}{|p{0.45\linewidth}|p{0.45\linewidth}|}
\hline
\textbf{Standard Dirac Theory (Abstract)} & \textbf{Hydrodynamic Analog (Concrete)} \\
\hline
\textbf{Mass ($m$):} A fundamental intrinsic constant. & \textbf{Mass:} An emergent resonance frequency ($\omega_0$) due to confinement. \\
\hline
\textbf{Negative Energy States:} Solutions with $E < -mc^2$, often requiring the "Dirac Sea" or hole theory interpretation. & \textbf{Inverted Anchoring Modes:} Mechanical modes where the topological winding is inverted due to excessive stress. \\
\hline
\textbf{Potential ($V$):} An external scalar/vector field modifying energy levels. & \textbf{Stress/Tension:} An external strain shifting the local stiffness of the medium. \\
\hline
\textbf{Klein Paradox ($R>1$):} Violation of single-particle probability; implies pair creation. & \textbf{Dielectric Breakdown:} Mechanical failure of the medium; implies generation of defect pairs to relax strain. \\
\hline
\end{tabular}
\end{table}

\subsection{Suggested Student Exercises}
To solidify understanding, instructors can assign the following problems based on the derivations in this text:

\begin{itemize}
    \item \textbf{Exercise 1 (Group Velocity vs. Current):} Using the dispersion relation in the supercritical region (Eq. 6), calculate the group velocity $v_g$ and demonstrate that it is positive (outgoing). Then, calculate the Dirac probability current $J$ for the antiparticle spinor and show it is negative. Discuss how this sign difference is crucial for obtaining $|R|^2 > 1$.
    
    \item \textbf{Exercise 2 (The Infinite Well Limit):} Following the logic in \textbf{Section \ref{sec:5}}, ask students to take the limit $V_0 \to \infty$ in the transmission coefficients found in standard textbooks (e.g., Greiner) and show that Klein tunneling vanishes. Students should be asked to interpret this physically: as the potential wall approaches infinite steepness and height without propagating into the medium, the boundary becomes infinitely rigid. Show how this extreme rigidity prevents the mechanical "breakdown" oscillation required to rip the vacuum, thereby suppressing pair production and recovering standard quantization.
    
    \item \textbf{Exercise 3 (The Schwinger Connection):} ask students to perform a dimensional analysis comparison between the pair production rate derived here (dependent on the stress gradient) and the exponential factor in the Schwinger limit $\exp(-\pi m^2/e\mathcal{E})$, identifying the analog of the electric field $\mathcal{E}$ (related to the strain gradient) in the elastic model.
\end{itemize}

\section{Conclusion}

In this work, we have investigated the Klein Paradox through the lens of a hydrodynamic analog model. By interpreting the Dirac equation as the effective field theory of a linear elastic medium, we demonstrated that the anomalous scattering coefficients arise from a mechanical instability analogous to dielectric breakdown under supercritical stress.

Within this framework, the ``negative energy'' modes are physically identified as excitations with inverted topological winding, and the reflection coefficient $R>1$ corresponds to the relaxation of vacuum strain via the generation of defect pairs. This derivation reproduces the pair production phenomenology of the Schwinger mechanism and the scattering results of Hansen and Ravndal \cite{Hansen1981}. Ultimately, this study reinforces the program of Analog Gravity by showing that complex quantum field theoretical processes can be rigorously modeled as the deterministic mechanics of a constrained continuum.

From a pedagogical perspective, this mechanical model offers a tangible alternative to the abstract algebraic derivations typically found in textbooks. By mapping the opaque concept of vacuum instability onto the intuitive physics of material breakdown, we provide students and instructors with a concrete visualization of the Dirac sea dynamics. We hope this framework serves as a valuable conceptual bridge in advanced courses —particularly those bridging relativistic equations with solid-state physics— facilitating a deeper intuition for the physical reality underlying the formalism of relativistic quantum mechanics in modern Dirac materials.

\appendix

\section{Emergence of Spinor Structure from Linearization Constraints}
\label{sec:app}

In this appendix, we demonstrate that the algebraic requirement of constructing a linear, first-order evolution equation from the second-order wave mechanics necessitates a multi-component state space structure isomorphic to the Dirac spinor. We show that the internal degrees of freedom required to stabilize the elastic defect map naturally onto the Clifford algebra of relativistic quantum mechanics.

\subsection{From Wave Mechanics to First-Order Dynamics}
The fundamental dynamics of a massive perturbation in a linear medium are governed by the second-order wave equation (Klein-Gordon form):
\begin{equation}
    \left( \frac{1}{c^2}\partial_t^2 - \nabla^2 + \frac{m^2 c^2}{\hbar^2} \right) \Psi = 0
\end{equation}
However, second-order equations allow for solutions with non-positive definite probability densities, describing transient excitations rather than persistent, conserved matter. To model a stable, conserved excitation (a "particle") with a conserved probability current, the equation of motion must be of \textit{first order} in time:
\begin{equation}
    i\hbar \frac{\partial \Psi}{\partial t} = \hat{H} \Psi
\end{equation}

\subsection{The Clifford Algebra Constraint}
Relativistic covariance requires that space and time be treated on equal footing. If the time derivative is linear, the spatial derivatives (momentum $\hat{\mathbf{p}}$) must also appear linearly in the Hamiltonian:
\begin{equation}
    \hat{H} = c \boldsymbol{\alpha} \cdot \hat{\mathbf{p}} + \beta mc^2
\end{equation}
For this linear ansatz to be consistent with the fundamental dispersion relation $E^2 = c^2 p^2 + m^2 c^4$ (derived from the resonance of the medium), the coefficients must satisfy the condition that squaring the operator recovers the scalar wave equation. This imposes the Clifford algebra:
\begin{equation}
    \{ \alpha_i, \alpha_j \} = 2\delta_{ij}, \quad \{ \alpha_i, \beta \} = 0, \quad \beta^2 = 1
\end{equation}
It is a well-known algebraic theorem that the smallest faithful representation of this algebra requires matrices of dimension $N=4$.

\subsection{Physical Identification of Components}
The mathematical requirement for a 4-component spinor maps naturally to the geometric degrees of freedom of a 1D elastic defect in 3D space. Specifically, the transverse vibrations in the 2D plane orthogonal to propagation require a 2-component complex representation (vibrational polarization or spin), while the topological boundary condition connecting the defect to the vacuum requires a binary orientation (topological anchoring or charge). Consequently, to satisfy the requirement of linear causal evolution within this elastic model, the effective field description requires a 4-component spinor representation, mirroring the Dirac field. 

In our defect model, this dimensionality corresponds to the physical degrees of freedom of the defect (anchored string)\footnote{This decomposition into two fundamental physical dualities is mathematically supported by the formalism of polar spinor decomposition, which shows that a Dirac spinor can be rigorously rewritten in terms of a real modulus, a chiral angle, and the parameters of a Lorentz transformation \cite{Fabbri2022deBroglie}. Our work provides a straightforward mechanical interpretation for these mathematical components}:
\begin{equation}
    \text{dim}(\Psi) = \text{dim}(V_{\text{spin}}) \times \text{dim}(V_{\text{anchor}}) = 2 \times 2 = 4
\end{equation}
The "antiparticle" solutions involved in the Klein paradox correspond to the subspace where the operator $\beta$ (acting on $V_{\text{anchor}}$) takes the eigenvalue $-1$. The paradox is thus resolved as a stress-induced transition between these topological sectors.

Physically, the topological index $\beta$ represents the orientation of the defect's anchoring relative to the medium's restorative field. If the defect's transverse displacement is aligned with the primary gradient of the vacuum's elastic restoring force, it occupies the $\beta = +1$ state (matter). Conversely, an excitation with $\beta = -1$ (antimatter) possesses an \textit{inverted} topological winding, meaning its phase response is exactly $\pi$ out of phase with the background restoring field. This physical inversion is what dictates the opposite flow of the probability current relative to the group velocity in the supercritical regime.

\begin{itemize}
  \item \textbf{The author declares that there are no conflict of interest.}
  \item \textbf{The author declares that no funding was provided for this research.}
\end{itemize}

%\bibliographystyle{apsrev4-1} 
%\bibliography{referencesKG} 

%merlin.mbs apsrev4-1.bst 2010-07-25 4.21a (PWD, AO, DPC) hacked
%Control: key (0)
%Control: author (72) initials jnrlst
%Control: editor formatted (1) identically to author
%Control: production of article title (-1) disabled
%Control: page (0) single
%Control: year (1) truncated
%Control: production of eprint (0) enabled
%

\end{document}